\documentclass{elsart}
\usepackage{graphicx}
\usepackage{amssymb}
\usepackage{amsfonts}
\usepackage[numbers,square]{natbib}
\begin{document}

\begin{frontmatter}

\title{\textbf{Economic dynamics with financial fragility and mean-field interaction: a model}}
\author[dipecon]{C. Di Guilmi \corauthref{cor}}, \author[dipecon]{M. Gallegati}, \author[ires]{S. Landini}

\address[dipecon]{Department of Economics, Universit\`a
 Politecnica delle Marche, Piazzale Martelli 8, 60121 Ancona, Italy}

\corauth[cor]{Corresponding author: Corrado Di Guilmi, Department of Economics, Universit\`a
 Politecnica delle Marche, Piazzale Martelli 8, 60121 Ancona,
 Italy. {\it Email address}: c.diguilmi@univpm.it.}

\address[ires]{IRES Piemonte - Istituto Ricerche Economiche e Sociali del Piemonte, Via Nizza 18, 10125 Turin, Italy}


\begin{abstract}
Following the statistical mechanics methodology, firstly introduced in
macroeconomics by Aoki \citep{Aoki96,Aoki02,Aoki06}, we provide some
insights to the well known works of \citeauthor{GS}
\cite{GS90,GS}. Specifically, we reach analytically a closed form
solution of their models overcoming the aggregation problem. The key
idea is to represent the economy as an evolving complex system, composed by
heterogeneous interacting agents, that can partitioned into a space of
macroscopic states. This meso level of aggregation permits to adopt mean
field interaction modeling and master equation techniques.
\end{abstract}

\begin{keyword}
Complex dynamics, master equation, financial fragility, mean field interaction, heterogeneity
\PACS 05.10.Gg, 89.65, 89.75.Fb 
\end{keyword}

\end{frontmatter}

\section{Introduction}

Aim of the present work is to replicate the mechanism of the models
introduced in \cite{GS} and \cite{JEBO} and to resettle them in a
dynamic stochastic framework, as
defined by Aoki \cite{Aoki96, Aoki02, Aoki06}. The
quantitative previsions of \cite{JEBO} (and of its developments
\citep{EmergentMacroeconomics}) replicate a number of stylized facts,
strengthening the idea that the economy would be better represented as
a complex dynamical system rather than a mere sum of identical and
perfect informed agents. Any attempt to resolve analytically this
kind of models must face the
aggregation issue, since the problem of how to sum up heterogeneous
and evolving agents cannot be dealt with the consuete tools of the
economist \cite{GallegatiPalestriniDelliGattiScalas}.
Masanao Aoki \citep{Aoki96, Aoki02, Aoki06} managed to remove the representative
agent hypothesis, introducing in economics the ground-breaking concept
of mean-field interaction, that makes feasible the analytical
aggregation of heterogeneous agents, replacing the unrealistic
mechanic determinism of  mainstream framework with a set stochastic
tools borrowed from statistical mechanics. Mean-field interaction can be defined as the average interaction model that substitutes all the relations among agents
that, otherwise, could not be analytically treated\cite{Opper01}. All
the agents are clusterized in a pre-defined set of states, basing on
one particular feature (the micro-state, e.g. the level of production for each firm) that determines the characteristics and the evolution
of the aggregate (the macro-state, e.g. the total level of
output). The accent is not on the single agent, but on the
number or proportion of agents that occupy a certain state at a
certain time. These levels are governed by a stochastic law, that
defines also the functional of the probability distributions of
aggregate variables and, if existing, their equilibrium
distributions.  The structure
of the work is the following: firstly, we specify the hypothesis for
the stochastic structure of the system (section \ref{sec:system} and for the firms that compose
it (section \ref{sec:firms})); then (section \ref{sec:stochastic}), we
develop the probabilistic framework, setting the dynamical instruments needed for
the aggregation.

\section{Structure of the system and definition of states}
\label{sec:system}

We set up a model in continuous
time for a system of heterogeneous and interacting agents, partitioned
into groups or states. In this paragraph we state the hypothesis that are at the root of this stochastic
dynamics, and in particular the definition and the structure of the
states. 
Our system is articulated in two states. This partition permits to isolate the effect
of bankruptcy costs on the
aggregate dynamics. Moreover, the application of recently proposed
methods for systems with higher order of states \citep{LaLama06} would entail a
too high price in terms of computational complication respect
to the expected improvement in the realism of the model.
Along time, a single firm can be in one of the two states, depending
on its financial soundness, proxied by the equity ratio (the ratio
among the net worth and the total assets). Therefore there are two
types of firms: the ``good'' firms, that have a high equity ratio, and
the ``bad'' firms, that have a low equity ratio that exposes them to
the risk of demise\footnote{To be complete, there is also a third type, the ``ugly'' firms, i.e. the
  failed ones, as explained in the following.}. System works in countinuos time \citep{HinichFosterWild2006}: $t \in\mathbb{T} \subset \mathbb{R}_+$.\\
The economy is populated by a fixed number of firms $N=N(t)\; : \; \forall t \in \mathbb{T}$, each indexed by $i$ for any given
time.\\
The system's vector of states $\underline{\omega}$ is identified by the financial
  condition of the firm:
  \small
\begin{displaymath}
    \underline{\omega}(t)= \left \{ \omega_i(t)=HV \left ( a_i(t) | \;
        \bar{a} \right ) \; \forall i \; \le N \right \} \ \ :\ \ HV(a_i(t)|\bar{a}) =\left \{ 
      \begin{array}{l}
1 \iff a_i(t)<\bar{a}\\
0 \iff a_i(t) \geq \bar{a}
      \end{array}
\right.
\end{displaymath}
  \normalsize
where $\bar{a}$ represents the threshold of equity ratio, that individuates firms that are in a critical financial
  situation, and, therefore, for which the probability of bankruptcy
   is bigger than $0$ and where $a_i(t)$ is the equity ratio for firm $i$ at time $t$.\\
In what follows we set $a_{1}(t)=x$ for firms which equity ratio is under the threshold and $a_{2}(t)=y$ for firms which equity ratio is over the threshold.\\
The cardinality of the $j$-th state, to say the number of firms in state $j=0,1$ is given by
  \begin{equation}
    \label{eq:occupation}
\left.
    \begin{array}{ll}
 card \; N_1(t)=\# \{\omega_i(t) =1 & \forall\ i \in I \}=N_1(t) \\ \\
 N_0(t)=N-N_1(t) & \\
    \end{array}
\right\}\Rightarrow
\underline{N}(t)=\left( N_0(t),N_1(t) \right )
  \end{equation}
  
 By assumption, the dynamics of the occupation number $N_j$
  follows a continuous time jump
  Markov process, defined over a state space $\Omega=(x,y)$ equipped
  with the following counting measure $N_{(.)}(.): 
   \Omega \times \mathbb{T}\to \mathbb{N}$, so that
  $(\Omega,N_{(.)}(.))$ is a countable sample space. In the
  following we indicate with $x$ the case of firms with equity ratio
  below the threshold $\bar{a}$ and with $y$ the alternative case:
  \begin{equation}
    \label{eq:defstate}
    \omega=x \Leftrightarrow \omega_i(t)=1 \ \lor \ \omega=y
    \Leftrightarrow \omega_i(t)=0 
  \end{equation}
such that $N_{(.)}(.)$ evaluates the cardinality of
  microstates: $N_{(\omega)}(t)=N_1(t) \iff \omega=x$ and
  $N_{(\omega)}=N_0(t) \iff \omega=y$. The relative frequency of firms is indicated in small
  letters: $n_k=N_k/N$. A-priori probability of $\omega=1$ is indicated by $\eta$:
  \begin{displaymath}
    p(\omega=x)=\eta \Leftrightarrow p(\omega=y)=1-\eta
  \end{displaymath}
A firm entries into the system in state $x$ and fails (and thus exits from
the system) only if it is in state $x$ (i. e. if its probability of bankruptcy is bigger than 0); in order to maintain constant the number of firms $N$ we assume
  that each bankrupted firm is immediately substituted by a new
  one. Therefore failures of firms do not modify the
  value of $N_1:=N(x)$. Firms move from $x$ to $y$ or vice versa according to the following
  transition rates:
\begin{equation}
  \label{eq:probjump}
\begin{array}{l}
  b(N_j)=r(N_j+1 \vert N_j)=\zeta \frac{N-N_j}{N}(1-\eta)\\
d(N_j)=l(N_j-1\vert N_j)=\iota \frac{N_j}{N}\eta
\end{array}
\end{equation}
where $\zeta$ is the transition probability from state $y$ to $x$ (firms whose
  financial position is deteriorated from a period to another, with equity ratio that
  becomes lower than $\bar{a}$) and $\iota$ the probability of the inverse transition (firms whose equity ratio improved
becoming bigger than $\bar{a}$). Having already indicated with $N^j$ the occupation number
of firms in state $j$, transition rates can be evaluated according to:
\begin{equation}
  \label{eq:transprobs}
  \begin{array}{l}
b(N_j)=r(N_j+1 \vert N_j)=\lambda(N-N_j): \lambda=\zeta (1-\eta)\\
d(N_j)=l(N_j-1 \vert N_j)=\gamma(N_j):\gamma=\iota \eta
  \end{array}
\end{equation}

In statistical mechanics terms, this kind of system
can be defined as a statistical ensemble with conservative cardinality,
described by a continuous time Markov process over a discrete state
space with the structure of a birth-death process. 



\section{The Firms}
\label{sec:firms}
\subsection{Hypothesis}
\label{sec:ipotesi}
The assumptions regarding firms are the same of the original models of \citeauthor{GS} \citep{GS, GS90}, as
modified by \cite{JEBO}, except for the following specifications. 
Respect to the theoretical construction
proposed by \cite{JEBO} this model does not consider any credit
market. There is no direct but a mean-field interaction. Differently from \cite{JEBO} the total number of firms
$N$ does not change over time. Firms are
identical within each state. The
  production function of a generic firm $i$ is:
    \begin{equation}
      \label{eq:fprod}
      q_i(t)=2(k_i(t))^{1/2}
    \end{equation}
where $k$ is the physical capital and $q$ is the physical output. It follows that the demand of capital function for
a single firms will be equal to:
  \begin{equation}
    \label{eq:capitaldemand}
      k_{i}(t)=\frac{1}{2}(q_i(t))^2
  \end{equation} 

The multiplicative shock price $\tilde{u_i(t)}$ has uniform distribution with support $[0.75;1.25]$ and $\mathbb{E}(\tilde{u})=1$. Given all the above specified hypothesis, the profit function for a
   generic firm $i$ can be expressed by:
   \begin{equation}
     \label{eq:profit}
     \pi_i(t)=P(t) \tilde{u}_i(t) q_i(t)-r_i(t) K_i(t)
   \end{equation}

 Once a firm get failed, it faces bankruptcy costs $C_i(t)
  \mu$ growing with the size of firm, proxied by the volume of its sales, and quantified by:
  \begin{equation}
    \label{eq:c}
 C_i(t)=c(P_i(t) q(t)_i)^2 =c(P(t) u_i(t) q_i(t))^2 \ \ :\ \  0<c<1 
  \end{equation}

\subsubsection{Transition probabilities}

A firm goes bankrupted if it "consumes" all its own capital, then $A_i\leq
  0$. So that, analogously with \cite{JEBO}, we can express the
    bankruptcy condition as a function of the price shock:
   \begin{displaymath}
       \tilde{u}(t') \leq \left( \frac{P(t)}{P(t')}\right)
       ( r K_i(t)/q_i(t)- a_1(t) \frac {K_i(t)}{P(t')q_i(t)} )\equiv \bar{u}_i(t')
  \end{displaymath}
 Substituting equation \ref{eq:capitaldemand} into the
above expression and, without loss of generality, normalizing reference price
$P(t)=P(t')$ to 1, it is possible to simplify the r.h.s. of
 the above equation: 

\begin{equation}
\bar{u}(t') \equiv \frac{q_1(t)}{2}(r-a_1(t) )   
\end{equation}

Recalling that the random variable $\tilde{u}$ has support $[0.75; 1.25]$, the critical thresholds of shock prices for having bankruptcy
will be:
\begin{equation}
\label{eq:baru}
\left \{
\begin{array}{ll}
\bar{u} = 0.75 \iff \tilde{u}_i(t)<0.75 \\
\bar{u} \in(0.75; 1.25) \iff \: 0.75 < \tilde{u}_i(t) < 1.25\\
\bar{u} = 1.25 \iff \tilde{u}_i(t)> 1.25
\end{array}
\right.
\end{equation}

Then, it is possible to indicate the probability $\mu(t)$ of failure for
a firm as a function of $\tilde{u}(t)$ :
\begin{equation}
 \label{bkrprob}
   \mu(t)=F(\tilde{u}(t))=p(\tilde{u}(t)\leq \bar{u}(t))=\frac{\bar{u}(t)-0.75}{0.5}=2\bar{u}(t)-1.5
\end{equation}

Equation (\ref{eq:bara}) permits to determine endogenously the threshold $\bar{a}$: indeed it can be interpreted as the minimum value of equity ratio which ensures
the surviving of the firm, i. e. for which the probability of
bankruptcy is equal to zero, and, therefore, it can be expressed as:
\begin{equation}
	\label{eq:bara}
	\bar{a}(t')=r-\frac{1.5}{q_1(t)}
\end{equation}

With an analogue procedure, we can specify the transition
  probabilities $\zeta$ and $\iota$ as dependent variables of the price shock
  $\tilde{u_i(t)}$. Indicating the critical values, respectively with $\bar{u}_{i,\zeta}(t)$ and
  $\bar{u}_{i,\iota}(t)$ we obtain:  
  \begin{displaymath}
    \begin{array}{l}
 \tilde{u}_i(t) \leq \frac{q_0(t)}{2}\left ( r+\bar{a}(t)-a_0(t) \right ) \equiv \bar{u}_{\zeta}(t) \\
 \tilde{u}_i(t) > \frac{q_1(t)}{2}\left (r+\bar{a}(t)-a_1(t) \right ) \equiv \bar{u}_{\iota}(t)
    \end{array}
  \end{displaymath}
since the thresholds, that for the \ref{bkrprob} is equal
to 0, here become equal to, respectively $(\bar{a}(t)-a_0(t))$ and
$(a_1(t)-\bar{a}(t))$. The range of variation of the two thresholds is
truncated as done in \ref{eq:baru}.
It is straightforward now to get the transition probability for each
state:
\begin{eqnarray}
  \label{eq:transratesu}
   \zeta(t)=p(\tilde{u}(t)\leq
   \bar{u}_{\zeta}(t))=2\bar{u}_{\zeta}(t)-1.5 \\
 \iota(t)=1-p(\tilde{u}(t)\leq
 \bar{u}_{\iota}(t))=-2\bar{u}_{\iota}(t)+2.5
\end{eqnarray}

\subsection{Firms object function.}
A firm decide the optimal quantity to produce in order to maximize its profit, using the information at
its disposal. Under the stated hypothesis the object function of a generic firm $i$ can
be then expressed as:
\begin{equation}
\max_{q_i(t)} F(q_i(t)):= \left \{ \mathbb{E} \left [ P(t) u_i(t')
    q_i(t) \right ] -rK_i(t)-C_i(t) \mu(t') \right \} 
\end{equation}
We suppose that firms take into consideration the present level of failure
probability, therefore that $\mu(t') =\mu(t)$, and that $P^e(t')=
P(t')=P(t)=1$ without loss of generality. 
The first order condition is:
\begin{displaymath}
 1-r q_i(t)-2cq_i(t)\mu(t)=0
\end{displaymath}
Consequently we get two different optimal levels of production for
firms in state $x$ and for
firms in state $y$, respectively:
\begin{eqnarray}
  \label{eq:qstar}
  q_{1}^{*}=(r+2c\mu)^{-1} \\
\nonumber q_{2}^{*}=r^{-1}
\end{eqnarray}
since $\mu=0$ for firms in state $y$.

\section{Stochastic inference}
\label{sec:stochastic}

\subsection{Master equation and stationary points}
In order to specify the dynamics of the joint probabilities, and, by this way, the
stochastic\footnote{Indeed, ``the master equation describes the time
  evolution of the probability distribution of states, not the time
  evolution of the states themselves'' \citep[page 7]{Aoki02}.}
evolution of the system, following \cite[chap. 3]{Aoki02},
\cite[page 252]{Landini} and \cite[page 62]{KuboTodaHashitume78}
we make use of the following master equation:
\begin{equation}
\label{eq:master}
\begin{array}{ll}
  \frac{dP(N_k,t)}{dt}&=b(N_{k-1})P(N_{k-1})+d(N_{k+1})P(N_{k+1})+\\&- \left \{ \left [ (b(N_k)+d(N_k))P(N_k) \right ] \right \} 
\end{array} 
\end{equation}
with boundary conditions:
\begin{equation}
\left\{\begin{array}{l}
    P(N,t)=b(N^1) P(N^1-1,t)+d(N) P(N,t)\\
P(0,t)=b(1)P(1,t)+d(0)P(0,t)
  \end{array} \right.
\end{equation}
These conditions ensure that the distributions functions consider only
consistent values, that is to say $N_1 \in [0;N]$ and, therefore, $n_1
\in [0;1]$. Since an analytical solution for master equations can be obtained
only under very specific conditions \citep{Risken89}, we solve it with an
approximation method, based on led and lag operators\footnote{For a detailed exposition see \cite{Aoki02} or \cite{Landini}.}. We assume that the number of
firms in state $x$ at a given moment is determined by its expected mean
($m$), the drift, and by an additive fluctuations component of order $N^{-1/2}$ around this value, that is to say the spread:

\begin{equation}\label{eq:drift}
N_1=Nm+\sqrt{N}s 
\end{equation}

Solving the modified master equation (see Appendix a) we find an explicit formulation for the macroscopic equation:
\begin{equation}
\label{eq:macroscopic}
\frac{d N_1}{dt}=-(\lambda+\gamma)N_1+\lambda N
\end{equation}
This equation describes the dynamics of the drift. It can be
interpretated as the long-run trend of the occupation number and,
then, keeping all the other relevant variables unchanged,
of the production. Now we can determine the stationary value of
$\mathbb{E}(n_k)=m$ and the consequent steady state equilibrium
of the economy, simply setting the r.h.s. of the
(\ref{eq:macroscopic}) to 0:
\begin{equation}
	\label{eq:stationarydrift}
	m^*=\frac{\lambda}{\lambda+\gamma}=n_1^*
\end{equation}

Once demonstrated the existence of a dynamical equilibrium point, a
deeper insight on system's dynamic is needed in
order to verify the convergence of the dynamics toward the
equilibrium distribution. The solution of the differential equation
(\ref{eq:macroscopic}) is:
\begin{equation}
  \label{eq:macrosolution}
  m(t)=m_0 e^{-(\lambda+\gamma)t}+\frac{\lambda}{\lambda+\gamma}
\end{equation}
that, setting an initial point $m_0$, yields to:
\begin{equation}
  \label{eq:macrostab}
  m(t)=m^*+(m_0-m^*)e^{-(\lambda+\gamma)t}
\end{equation}
that verifies the convergence and the stability of the equilibrium
since the second term goes to 0 as $t  \to \infty$.

\subsection {Aggregate output}

The aggregate output of the system will be:
\begin{equation}
  \label{eq:Y}
   Y(t)=\frac{N_1}{r+2c\mu(t)}+\frac{N_0}{r}
 \end{equation}

Keeping constant all the other variables, dynamic
fluctuations in the level of $Y(t)$ are due to the changing in the
levels of $N_1$ and $N_0$. 
Once quantified the equilibrium distribution of the drift, we
are able now to obtain the steady state value of aggregate production:
\begin{equation}
  \label{eq:stationaryY}
  Y^e=N \left [
    \frac{1}{r}-\frac{\lambda}{\lambda+\gamma}\frac{2c\mu}{r(r+2c\mu)}
  \right ]= N \left [\frac{1}{r}-\frac{\lambda}{\lambda+\gamma} \left
      ( q_1-q_0 \right ) \right ]
\end{equation}

As can be easily seen the factor inside the brackets of the (\ref{eq:stationaryY}) is
equal to 
\begin{displaymath}
  \left [ y_0-m^*(y_1-y_0) \right ]
\end{displaymath}

 that testifies the consistency of the model. Since the aggregate
 production function depends on $m$, also its dynamic will be convergent
 to a stationary level. 
Dynamics comes out to be dependent on the transition rates
$\lambda$ and $\gamma$ and on the differences in firms level of
production. The dynamics of these factors are studied in the next section.

\subsection{Equilibrium distribution and critical points}
\label{par:potential}
Equating the master equation to 0, it is possible to obtain the
Kolmogorov condition that equates the
probability fluxes entering in a state with the fluxes coming out from
that state. Formally, using equations \ref{eq:transprobs} and
\ref{eq:macroscopic}:
\begin{equation}
  \label{eq:macroscopic2}
  \dot{m}=\zeta \eta(m)- \left [ \iota (1-\eta(m))+\zeta
    \eta(m)\right ]m
\end{equation}
Setting $\dot{m}$ to 0 and rearranging, the stationary configuration
of the system can be expressed as:
\begin{equation}
  \label{eq:stationary2}
  \dot{m}=0 \Rightarrow \frac{\eta(m^*)}{1-\eta(m^*)}=\frac{\iota m^*}{\zeta(1-m^*)}
\end{equation}
Being in Markovian space, we can make use of Brook's lemma
\citep{Brook64} that defines
local characteristic for this kind of chains:
\begin{equation}
  \label{eq:brook}
  P^e(N_k)=P^e(N_0) \left ( \frac {\iota}{\zeta} \right )  {N \choose
    N_k} \prod^k_{j=1} \frac{\eta(N-N_j)}{(1-\eta(N_j))}
\end{equation}
By means of Hammersley and Clifford theorem (see \cite{HammersleyClifford71} as demonstrated in
\cite{Clifford90}), the stationary probability of the markovian process for $N_j$, when detailed balance holds, can be
expressed by:
\begin{equation}
  P^e(N_j) \propto Z^{-1}e^{-\beta N U(N_j)}
\end{equation}
where $U(N^j)$ is the {\it Gibbs
  potential} \citep{Woess96}. The parameter $\beta$ may be interpreted as an inverse measure of the uncertainty
of the system. The above formulation allows us to express explicitly the values of
the a-priori probabilities:
\begin{eqnarray}
  \label{eq:probhc}
  \eta(N_j)=N^{-1}e^{\beta g(N_j)}\\
1-\eta(N_j)=N^{-1}e^{-\beta g(N_j)}
\end{eqnarray}
so that:
\begin{equation}
  e^{\beta g(N_j)}+ e^{-\beta g(N_j)}=N
\end{equation}
It is easy to verify that large values of $\beta$ associated with
positive values of $g(N_j)$ cause $\eta(N_j)$ to be larger than
$1-\eta(N_j)$, making transition from state $y$ to state $x$ more
likely to occur than the opposite one. On the other hand, values of
$\beta$ close to 0, make $\eta(N_j)$ close to 0.5.
To get a deeper insight on the meanings of $\beta$ let us carry out
the potential equation. In binary models and for great $N$, the equation of the
potential is:
\begin{displaymath}
  U(N^j)=-2\int^{N^j}_0 g(N_j)dy-\frac{1}{\beta} H(\underline{N})
\end{displaymath}
where $H(\underline{N})$ is the Shannon entropy for the vector of
occupation numbers. $g(N_j)$ is a function
that evaluates the relative difference in the outcome as a function of
$N_j$. In intuitive terms, to individuate the stationary points of
probability dynamics we need to individuate its peak. $\beta$ is an inverse multiplicative factor for entropy: this
implies then that, for very large values of $\beta$, the entropy
component does not play any role. On the contrary, as $\beta$
approaches 0, the weight of the entropy component grows. In terms of
the present model, a relative high value of $\beta$ means that the
uncertainty in the system is low, with few firms exposed at bankruptcy
risk and, due to this, the bankruptcy probability $\mu$ plays a
negligible role \citep[pp. 55 and
following]{Aoki96}. Therefore for values of $\beta$ around 0, and a
more relevant volatility in the system, in order
to individuate the peak of probability dynamic we need to find the
local minimum of the potential. \cite{Aoki02} shows that the points in
which the potential is minimized are also the critical point of the
aggregate dynamics of $p_j$. Deriving the potential respect to $N^j$:
\begin{equation}
\label{eq:derpotential}
  g(N^j)=-\frac{1}{2\beta}\frac{dH(N_j)}{dN_j}=-\frac{1}{2\beta}\ln
  \left (  \frac{N_j}{N-N_j} \right )
\end{equation}
Solving the following MaxEnt problem \citep{Jeynes57}:

\begin{equation}
  \label{eq:constraints}
\left \{
\begin{array}{lc}
 \max  H(N_1,N_0)=-N_1(t) \ln(N_1(t))-N_0(t)\ln(N_0(t)) &\; s.t.\\ 
N_1(t)+N_0(t)=N  \\
N_1(t)y_1(t)+N_0(t)y_0(t)=Y(t)  \\
\end{array}
\right.
\end{equation}
 with suitable Lagrange multipliers equal
 to, respectively, $\lambda_1=1-\alpha$ and $\lambda_2=-\beta$ we get a
 closed solution for $\beta$:
\begin{equation}
  \label{eq:betasenza}
  \beta(t)= \ln  \left( - \frac {y_1(t)-\bar{y(t)}}{ y_0(t)-\bar{y(t)}} \right)  \left( y_1(t)-y_0(t) \right) ^{-1}
\end{equation}
Setting $U'=0$ and using equation (\ref{eq:betasenza}), we get an explicit formulation for $g(N_1)$ in
stationary condition:
\begin{equation}
  \label{eq:gn1}
  g(N_1)=\frac{y_0-y_1}{2}
\end{equation}
that quantifies the mean difference (for states) of the outcome. \\
From equation \ref{eq:derpotential} it follows that the
point of local minimal of the potential is given by:
\begin{equation}
  \label{eq:minpot}
  U'=0 \Rightarrow e^{2\beta g(N_j)} =\frac{N_j}{N-N_j}
\end{equation}
that, if the rates of entries and exits are equated,
(i. e. if $\iota=\zeta$), reproduces exactly the \ref{eq:stationary2}.
Therefore, making use of the \ref{eq:probhc} we can write:
\begin{equation}
  \label{eq:stationarygibbs}
  N_j=\frac{ e^{2\beta g(N_j)}}{ e^{2\beta g(N_j)}+ e^{-2\beta g(N_j)}}
\end{equation}
that is the maximum likelihood estimation of the Gibbs distribution of
the firms in state $N_j$.
Let us analyze the different behavior of the
stationary distribution for different values of $\beta$. \\
\fbox{$\beta \rightarrow \infty$} Given equation \ref{eq:derpotential}, the critical points in which the potential is minimized
  are also the zeros of the function $g(N_j)$:
  \begin{equation}
    \label{eq:betainf}
    U'(m^*)=-2g(m^*)=0
  \end{equation}
This suggests that $\beta$ may be interpreted as an inverse index of
 uncertainty\citep{Aoki96}. Considering equation \ref{eq:gn1}:
 \begin{displaymath}
   g(N_j)=0 \Leftrightarrow y_0-y_1=0
 \end{displaymath}
Under the specified conditions, there is no uncertainty in the system, since no firm
can go bankrupted and, therefore, the level of production is
unambiguously determined. Indeed the
value of $\beta$ can go to infinity if $N_0 \to N$ or if $\mu \to 0$,
since both situation imply a convergence among the different targets of
production at micro level and, then, a minimum degree of uncertainty in
the system.\\
\fbox{$\beta \rightarrow 0$} In this second case, in order to individuate the
  critical points of the dynamics, a further deepening is
  needed.  $\beta$ can go to $0$ if and only if
  $\frac{N_j}{N-N_j} \to 1$, that is to say, if the system is populated,
  quite in the same proportion, of firms in state $x$ and of firms in
  state $y$. But this is not informative about the behavior of the
  $g(N)$ since the first factor of the equation \ref{eq:derpotential}
  goes to infinity (given that $\beta \to 0$) while the second goes to
  $0$. The Cox and Miller hazard function \citep{CoxMiller96}, with a density
function analogous to the one of equal to a standard Brownian motion's
first-passage\citep{GrimmettStirzaker92}, in terms of $m$ can be
expressed as:
\begin{equation}
  \label{eq:hazard1}
  F(m)=\left [ 1+e^{-2\beta m} \right ]^{-1} \Rightarrow
  h(m)=\frac{2 \beta}{1+e^{-2\beta m}}
\end{equation}

Now, we have to calculate the probability that a firm passes from
a state to another in response to a small variation in the difference
of relative production, conditional on the current difference among
$y_0$ and $y_1$ (quantified by $g(m)$, from equation
\ref{eq:gn1}). We can then rewrite the conditional
hazard function in the following way:

\begin{equation}
  \label{eq:hazardgm}
  P(v \leq y_1-y_0 \vert m^*)= \left [ 1+e^{-2 \beta g(m^*)} \right ]^{-1}
\end{equation}
and then:
\begin{equation}
  \label{eq:hazard2}
  h(y_1-y_0 \vert m^*)=\frac{2\beta \eta(m^*)}{1+e^{-2\beta g(m^*)}}
\end{equation}
Supposing that $\eta(m^*)=m^*$, we finally obtain:
\begin{equation}
  \label{eq:hazard3}
 h(y_1-y_0 \vert m^*)= 2 \beta m^* 
\end{equation}
Therefore, we may conclude that for values of $\beta$ close to 0, the
critical point of probability dynamics, here found by minimizing the
potential, is a value of $m^*$ approximately equal to $\beta$
itself. In other words, $\beta$ may be considered the as the
conditional hazard rate in the range where $\beta$ is small. The
potential then is minimized for a fraction $m^*$ of firms in state $x$
when the value of the conditional hazard function is approximately
equal to $\beta$. 

\section{Concluding remarks}
In this work we apply Aoki's methodology to the model
presented in \cite{GS} and modified, introducing interacting
heterogeneous agents, in \cite{JEBO}. The adoption of the statistical
mechanical approach, based on mean field interaction and aggregate stochastic
dynamic analysis, leads to a stable analytical solution, without recurring to computer simulations. We
start from very general conditions and hypothesis, embodied in a
stochastic framework, and, finally, we obtain an explicit macroscopic equation that describes the
evolution of the system and its long term equilibrium solution. The
dynamics is analyzed by means of master equation
solution techniques enriched by the use of MaxEnt and hazard function analysis.
\\

\appendix \textbf{Appendix A}
\label{appendixa}
\\
The master equation has now to be expressed as $\dot{Q}(s)$, a function
of $s$, and then becomes:
\begin{equation}
  \label{eq:mewithdrift}
  \dot{P}(N_k)=\frac{\partial Q}{\partial t}-\frac{d s}{d t}
  \frac{\partial Q}{\partial s}=\dot{Q}(s)
\end{equation}
with transition rates reformulated in the following way:
\begin{eqnarray}
  \label{eq:trdrift}
  b(s)=\lambda \left [N-Nm-\sqrt{N}s \right ]\\
d(s)=\gamma \left [ Nm+\sqrt{N}s \right ]
\end{eqnarray}
 Since
 \begin{equation}
\label{eq:dsdt}
   \frac{d s}{d t}=-N^{1/2} \frac{d m}{d t}
 \end{equation}
equation (\ref{eq:mewithdrift}) can be expressed as:
\begin{equation}
  \label{eq:mewithdrift2}
  \dot{Q}(s)=\frac{\partial Q}{\partial t}-N^{1/2}\frac{\partial Q}{\partial s}\dot{m}
\end{equation}

We rewrite again now the master equation (\ref{eq:master}) and the
transition rates by means of
lead and lag operators. The use of these operators allows to express the master equation in a more treatable
form, making the two probability flows (in and out)
homogeneous. Specifically the transition
probabilities (\ref{eq:transprobs}) become:
\begin{eqnarray}
\label{eq:transprobll}
  L[d(N_k)P(N_k,t)]=d(N_{k+1})P(N_{k+1})\\
L^{-1}[b(N_k)P(N_k,t)]=d(N_{k-1})P(N_{k-1})
\end{eqnarray}
so that the master equation will be expressed in this way:

\begin{equation}
  \label{eq:mell}
  \dot{Q}(s)=(L-1)[d(s)Q(s)]+(L^{-1}-1)[d(s)Q(s)]
\end{equation}

Using the modified transition rates (\ref{eq:transprobll}) and expanding
the so obtained master equation in
inverse powers of $s$ to the second order we get:
\begin{equation}
\begin{array}{l}
  \label{eq:mell2}
  
N^{-1} \frac {\partial Q}{\partial
  \tau}-N^{-1/2}\frac{dm}{d\tau}\frac{\partial Q}{\partial s}=\\
N^{-1/2} \left ( \frac{\partial}{\partial s} \right ) [d(s)Q(s)] +
N^{-1} \frac{1}{2} \left ( \frac{\partial}{\partial s} \right )^2[d(s)Q(s)]+\\
-N^-1/2 \left ( \frac{\partial}{\partial s} \right )[b(s)Q(s)] +
N^{-1}\frac{1}{2}\left ( \frac{\partial}{\partial s} \right )^2[b(s)Q(s)] + . . .\\
= N^{-1/2} \left ( \frac{\partial}{\partial s} \right )[(d(s) -
b(s))Q(s)] + N^{-1}\frac{1}{2}\left ( \frac{\partial}{\partial s} \right )[(b(s) + d(s))Q(s)] + . . .
\end{array}
\end{equation}
At this point, in order to
match the component of the same orders of powers of $N$ between and
equations (\ref{eq:mewithdrift}) and (\ref{eq:mell2}), we need to rescale the variable $\tau=t N$.
Knowing that:
\begin{displaymath}
  \begin{array}{l}
d(s)-b(s)=(\lambda+\gamma)(Nm+\sqrt{N}s)-\lambda
N=(\lambda+\gamma)N_k-\lambda N\\
d(s)+b(s)=(\lambda-\gamma)(Nm+\sqrt{N}s)+\lambda N=(\lambda-\gamma)N_k+\lambda N
  \end{array}
\end{displaymath}

and taking the derivatives, up to the second order, it is possible to
obtain what Aoki defines as {\it diffusion approximation}
\cite{Aoki02}:
\begin{equation}
\label{eq:diffusapp}
\begin{array}{l}
N^{-1} \frac{\partial Q}{\partial \tau}-
N^{-1/2}\frac{dm}{d \tau}\frac{\partial Q}{\partial s}=\\
(\lambda+\gamma)Q(s)+N^{-1/2}(d(s)-b(s))\left ( \frac{\partial}{\partial s} \right )Q(s)+N^{-1} \frac{1}{2}(b(s)+d(s))\left ( \frac{\partial}{\partial s} \right )Q(s)
\end{array}
\end{equation}
Equating the terms of order $N^{-1/2}$ for the polynomial identity
principle we get:
\begin{displaymath}
N^{-1/2} \frac{dm}{dt}\frac{\partial Q}{\partial s}=-N^{-1/2}(b(s)- d(s))\left ( \frac{\partial}{\partial s} \right )Q(s)
\end{displaymath}

\end{document}